\documentclass[9pt,twocolumn,twoside]{pnas-new}

\templatetype{pnasresearcharticle} 

\setboolean{displaywatermark}{false}
\usepackage[caption=false]{subfig}
\usepackage{amsmath}
\usepackage{bm}

\title{Quasicrystalline 30\textdegree~Twisted Bilayer Graphene as an Incommensurate Superlattice with Strong Interlayer Coupling}

\author[a]{Wei Yao}
\author[a]{Eryin Wang} 
\author[a]{Changhua Bao}
\author[b]{Yiou Zhang}
\author[a]{Kenan Zhang}
\author[b]{Kejie Bao}
\author[b]{Chun Kai Chan}
\author[c]{Chaoyu Chen}
\author[c]{Jose Avila}
\author[c,d]{Maria C. Asensio}
\author[b,1]{Junyi Zhu}
\author[a,e,1]{Shuyun Zhou}

\affil[a]{State Key Laboratory of Low Dimensional Quantum Physics and Department of Physics, Tsinghua University, Beijing 100084, China}
\affil[b]{Department of Physics, The Chinese University of Hong Kong, Hong Kong, China}
\affil[c]{Synchrotron SOLEIL, L'Orme des Merisiers, Saint Aubin-BP 48, 91192 Gif sur Yvette Cedex, France}
\affil[d]{Universit\'e Paris-Saclay, L'Orme des Merisiers, Saint Aubin-BP 48, 91192 Gif sur Yvette Cedex, France}
\affil[e]{Collaborative Innovation Center of Quantum Matter, Beijing, P.R. China}

\leadauthor{Yao} 

\significancestatement{Interlayer interaction in van der Waals heterostructures could induce many exotic phenomena. Unlike commensurate heterostructures, incommensurate ones have often been neglected, due to the rarity in nature and the assumption of suppressed coherent interlayer movement of electrons. How the interlayer interaction affects the electronic structure of such non-periodic heterostructure is a fundamental question. In this work, by successfully growing 30\textdegree-tBLG as an example for incommensurate ``heterostructure'' with quasicrystalline order, we report the emergence of novel mirrored Dirac cones. We identify that these mirrored Dirac cones are a consequence of the interlayer interaction, showing its importance in the incommensurate structure which has been overlooked before. These results broaden the application range of van der Waals heterostructure for future electronic devices.}

\authorcontributions{Author contributions: S.Z. conceived the research project. W.Y. prepared the graphene sample. W.Y. and K.Z. performed the Raman measurements. W.Y. and E.W. performed the conventional ARPES measurements. E.W. and C.B. performed Nano-ARPES measurements and analyzed the ARPES data with assistance from C.C., J.A. and M.C.A. Y.Z., K.B., C.K.C. and J.Z. performed the theoretical calculation. W.Y., J.Z. and S.Z. wrote the manuscript, and all authors commented on the manuscript.}
\authordeclaration{The authors declare no conflict of interest.}
\correspondingauthor{\textsuperscript{1}To whom correspondence should be addressed. E-mail: \href{mailto:jyzhu@phy.cuhk.edu.hk}{jyzhu@phy.cuhk.edu.hk} and \href{mailto:syzhou@mail.tsinghua.edu.cn}{syzhou@mail.tsinghua.edu.cn}.}


\begin{abstract}
The interlayer coupling can be used to engineer the electronic structure of van der Waals heterostructures (superlattices) to obtain properties that are not possible in a single material. So far research in heterostructures has been focused on \textit{commensurate} superlattices with a long-ranged Moir\'{e} period. \textit{Incommensurate} heterostructures with rotational symmetry but not translational symmetry (in analogy to quasicrystals) are not only rare in nature, but also the interlayer interaction has often been assumed to be negligible due to the lack of phase coherence. Here we report the successful growth of quasicrystalline 30\textdegree~twisted bilayer graphene (30\textdegree-tBLG) which is stabilized by the Pt(111) substrate, and reveal its electronic structure. The 30\textdegree -tBLG is confirmed by low energy electron diffraction and the intervalley double-resonance Raman mode at 1383 cm$^{-1}$. Moreover, the emergence of mirrored Dirac cones inside the Brillouin zone of each graphene layer and a gap opening at the zone boundary suggest that these two graphene layers are coupled via a generalized Umklapp scattering mechanism, i.e. scattering of Dirac cone in one graphene layer by the reciprocal lattice vector of the other graphene layer. Our work highlights the important role of interlayer coupling in incommensurate quasicrystalline superlattices, thereby extending band structure engineering to incommensurate superstructures.
\end{abstract}

\dates{This manuscript was compiled on \today}
\doi{\url{www.pnas.org/cgi/doi/10.1073/pnas.XXXXXXXXXX}}

\begin{document}

\verticaladjustment{-2pt}

\maketitle
\thispagestyle{firststyle}
\ifthenelse{\boolean{shortarticle}}{\ifthenelse{\boolean{singlecolumn}}{\abscontentformatted}{\abscontent}}{}



\begin{figure}
	\subfloat{\label{fig1a}}
	\subfloat{\label{fig1b}}
	\subfloat{\label{fig1c}}
	\centering
	\includegraphics[width=8.5 cm] {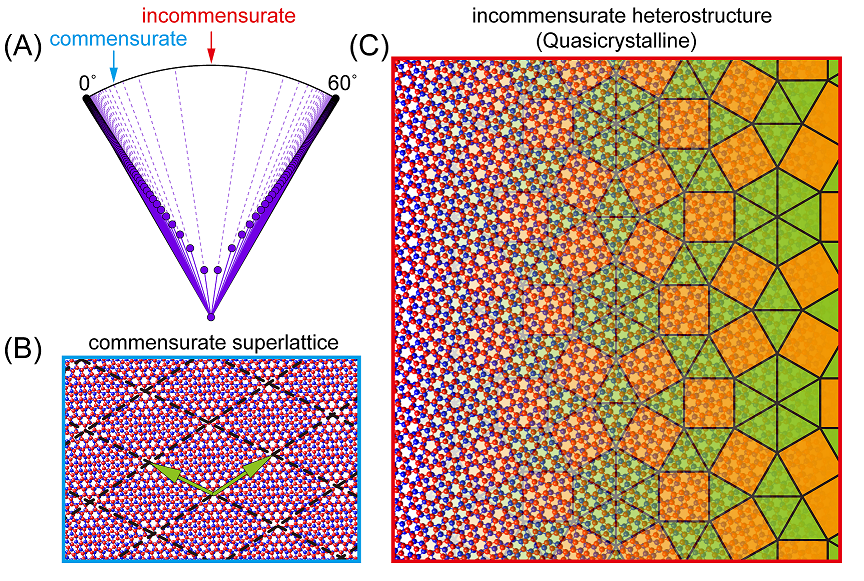}
	\caption{\textbf{Commensurate and incommensurate twisted bilayer graphene.}  \textbf{(A)}, The distribution of all possible twisting angles $\theta_t$ for forming commensurate tBLG when $q=1$ in Eq.~(1). The Moir\'{e} period at corresponding twisting angles is represented by the length of solid line (with logarithmic scale). \textbf{(B)}, Commensurate tBLG with 7.34\textdegree~twisting angle (indicated by the blue arrow in (A)), showing Moir\'{e} pattern with periodic translational symmetry. Green arrows are the Moir\'{e} reciprocal lattice vectors. \textbf{(C)}, Incommensurate 30\textdegree-tBLG without any long-ranged period (indicated by red arrow in (A)), showing patterns of dodecagonal quasicrystal.}
\end{figure}

\dropcap{I}n van der Waals heterostructures \cite{Geim2013vdW}, a long-ranged commensurate Moir\'{e} superlattice \cite{HoneNano2010, LeRoyNatureMater11, KimNature13, GBNgap,GeimNaturePhys2014, WangNP16} forms only when satisfying $m\bm{a}_1^{\rm top}$+$n\bm{a}_2^{\rm top}$=$m^\prime$$\bm{a}_1^{\rm bottom}$+$n^\prime$$\bm{a}_2^{\rm bottom}$ \cite{KoshinoPRB12}, where $\bm{a}_{1,2}^{\rm top}$ and $\bm{a}_{1,2}^{\rm bottom}$ are primitive lattice vectors for the top and bottom layers respectively, and $m$, $n$, $m^\prime$, $n^\prime$ are integers.  Twisted bilayer graphene (tBLG) can be viewed as a simple version of ``hetero''-structure with $\bm{a}_{1,2}^{\rm top}$ rotated by a twisting angle $\theta_t$ with respect to $\bm{a}_{1,2}^{\rm bottom}$. In tBLG, commensuration occurs when $\theta_t$ satisfies \cite{Mele2010prb,Neto2012prb}
\begin{equation}
\label{eq1}
\cos{\theta_t}=\frac{3p^2+3pq+q^2/2}{3p^2+3pq+q^2}
\end{equation}
where $p$ and $q$ are two coprime positive integers. Figure~\ref{fig1a} plots the possible twisting angles for forming commensurate tBLG for $q=1$. It is clear that commensuration easily forms around 0\textdegree~while for larger twist angles, especially around 30\textdegree, incommensurate superlattice is more common \cite{Novoselov2014NP}. In contrast to commensurate superlattice with a long-ranged period (see example in Fig.~\ref{fig1b}), incommensurate superlattice lacks a long-ranged translational symmetry in real space while preserving the rotational symmetry,  e.g. 30\textdegree-tBLG (Fig.~\ref{fig1c}) is similar to quasicrystals with a classic dodecagonal pattern \cite{Koren2016prb}.

Although the electronic structures of commensurate heterostructures have been investigated in recent years \cite{Andrei2010np, Andrei2011prl,Eli2012prl,Barinov2015SciRep,Vignaud2016SciRep}, research on incommensurate heterostructures remains limited for two reasons. On one hand, incommensurate heterostructures are difficult to be stabilized and thus they are quite rare under natural growth conditions. On the other hand, it is usually assumed that the interlayer interaction is suppressed due to the lack of phase coherence. For example, incommensurate tBLG has often been taken as a trivial combination of two non-interacting graphene layers \cite{deHeerPRL,HaasPRL08,JSKim2013prl}. How the electronic structure is modified in such incommensurate heterostructures, with symmetry analogous to quasicrystalline order, is a fundamental question. Here by growing epitaxial 30\textdegree-tBLG successfully on Pt(111) substrate and revealing its electronic structure using angle-resolved photoemission spectroscopy (ARPES) and NanoARPES, we show that  the interlayer coupling can modify the electronic structure significantly, leading to mirrored Dirac cones.  Additionally, to understand the intriguing stability that is against common belief on bilayer graphene, we performed first-principles calculations and found that such incommensurate  30\textdegree-tBLG heterostructure is stabilized under appropriate substrate conditions.

\section*{Results}

\subsection*{Growth of 30\textdegree-tBLG}

The 30\textdegree-tBLG sample was grown on Pt(111) substrate by carbon segregation from the bulk substrate \cite{SutterPRB, YW2015prb}.  Figure \ref{fig2a} shows the low energy electron diffraction (LEED) of bottom monolayer graphene at 30\textdegree~azimuthal orientation (blue arrow) relative to the substrate.  Our previous work shows that distinguished from other graphene/Pt with commensurate (e.g. 2$\times$2, 3$\times$3)  Moir\'{e} superstructures \cite{SutterPRB}, the interface between the 30\textdegree~monolayer graphene and the Pt(111) substrate is incommensurate without forming any Moir\'{e} pattern, leading to nearly free-standing monolayer graphene  \cite{YW2015prb}. Further increasing the annealing temperature and time can lead to thicker graphene sample with a new set of diffraction peaks emerging at 0\textdegree~orientation (red arrow in Fig.~\ref{fig2b}) coexisting with those at 30\textdegree, and such graphene sample is the focus of current work. The lattice constants extracted from LEED show that there is negligible strain (<0.2$\%$, see Fig.~S1 in SI Appendix for details) between the 30\textdegree~and 0\textdegree~graphene layers, and the absence of additional reconstructed diffraction spots in the LEED pattern further supports that they do not form commensurate superlattice. If these diffraction peaks come from the same graphene domains, this would imply that 0\textdegree~graphene is stacked on top of the bottom 30\textdegree~graphene layer, namely, a 30\textdegree-tBLG is formed. In the following, we will provide direct experimental evidence for the conjectured incommensurate 30\textdegree-tBLG from Raman spectroscopy and reveal its electronic structure from ARPES measurements.

\subsection*{Intervalley double-resonance Raman mode}

\begin{figure*}
	\subfloat{\label{fig2a}}
	\subfloat{\label{fig2b}}
	\subfloat{\label{fig2c}}
	\subfloat{\label{fig2d}}
	\subfloat{\label{fig2e}}
	\subfloat{\label{fig2f}}
	\subfloat{\label{fig2g}}
	\subfloat{\label{fig2h}}
	\subfloat{\label{fig2i}}
	\subfloat{\label{fig2j}}
	\centering
	\includegraphics[width=17.4 cm] {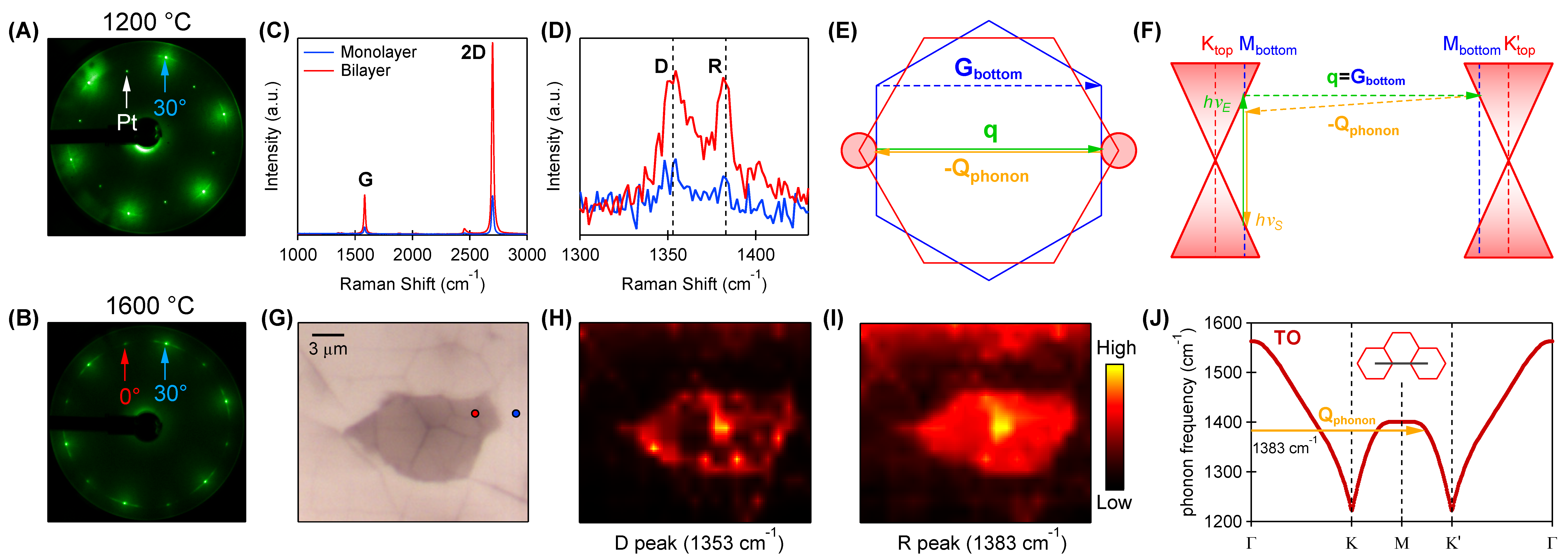}
	\caption{\textbf{Observation of double-resonance Raman mode in 30\textdegree-tBLG.}  \textbf{(A)}, The LEED pattern for the graphene sample after annealing at 1200 \textcelsius. \textbf{(B)}, The LEED pattern for the graphene sample at higher annealing temperature of 1600 \textcelsius. \textbf{(C)}, Measured Raman spectrum for monolayer and bilayer region respectively, covering the range of G peak and the 2D peak. \textbf{(D)}, Zoom in of (C) for range from 1300 to 1430 cm$^{-1}$. \textbf{(E)}, The geometry for electron momentum transferring in DR Raman process of the observed R mode. \textbf{(F)}, The schematic drawing for the DR process of R mode in graphene band structure (not in scale). Electron at one Dirac cone from the top layer is photo-excited to the conduction band, scattered by one reciprocal lattice vector of the bottom layer $\bm{q}=\bm{G}_{\rm bottom}$, and subsequently scattered back by phonon with vector $\bm{Q}_{\rm phonon}$. \textbf{(G)}, The optical image for the measured area. The red and blue dots indicate the measuring positions of spectra in (C) and (D). \textbf{(H)}, The Raman map for the D peak, integrating intensity from 1343 to 1363 cm$^{-1}$. \textbf{(I)}, The Raman map for the R peak, integrating intensity from 1373 to 1393 cm$^{-1}$. \textbf{(J)}, The phonon spectrum of the transversal optical (TO) mode along high symmetric direction, taken from Ref.~\cite{CarozoRaman}. The arrow indicates the momentum of transferring R mode phonon. The inset shows the high-symmetry path in $k$-space for the phonon spectrum. The left hexagon is the first Brillouin zone.}
\end{figure*}

Raman spectroscopy is a powerful tool for characterizing the vibrational mode in graphene \cite{FerrariRaman} and can provide direct information about the sample thickness and stacking. Figure \ref{fig2g} shows a typical optical image of the as-grown sample. The optical image shows strong intensity contrast, with a darker region of $\approx$ 10 $\mu$m overlapping with the brighter region. The Raman spectra in Fig.~\ref{fig2c} show stronger intensity for the darker region (red curve) than the brighter region (blue curve), suggesting that the darker region is thicker. The spectra for both regions show characteristic features of monolayer graphene \cite{FerrariRaman}, in which the 2D mode shows a single Lorentzian peak with stronger intensity than the G mode. This suggests that the top and bottom flakes are both monolayer graphene, and the thicker region is not a Bernal (AB stacking) bilayer graphene, but a bilayer graphene with a large twisting angle instead \cite{twistRaman}. More importantly, when zooming in the spectra between 1300 to 1430 cm$^{-1}$,  the bilayer region shows two peaks centered at 1353 cm$^{-1}$ and 1383 cm$^{-1}$ respectively (Fig.~\ref{fig2d}). The peak at 1353 cm$^{-1}$ is the D mode of graphene which is caused by the limited size or defects \cite{FerrariRaman}. The Raman mapping for this peak (Fig.~\ref{fig2h}) shows that it only appears at the edges of the bilayer region or defect centers, consistent with the nature of the D peak. In contrast, the peak at 1383 cm$^{-1}$ (``R'' peak) is observed in the whole bilayer graphene region (Fig.~\ref{fig2i}), indicating that it is intrinsic to the twisted bilayer graphene. 

Since the energy of the R mode is close to that of the D mode, the R mode likely has similar origin as the D mode: intervalley double-resonance (DR) Raman process \cite{DoubleResonance,CarozoRaman}. The key process for the intervalley DR Raman process involves a special scattering process (by defect, phonon or Moir\'{e} pattern etc.) in which photoexcited electrons at one valley are scattered to another valley by a non-zero momentum transfer $\bm{q}$ and subsequently decay back to the original valley by emitting a phonon with wave vector $\bm{Q_{\rm phonon}}$=$\bm{q}$ (see Fig.~\ref{fig2f}). While D mode is induced by defect scattering between two neighboring Dirac cones, instead, we claim that the R mode in our case is caused by scattering between two opposite Dirac cones which are connected by one reciprocal lattice vector of the bottom graphene layer with $\bm{q}$=$\bm{G_{\rm bottom}}$ (see Fig.~\ref{fig2e} and \ref{fig2f}). This is verified by observing a phonon mode with matching momentum and energy in the transverse optical (TO) phonon spectrum along $\Gamma$-K-M-K$^\prime$-$\Gamma$ direction (see Fig.~\ref{fig2j}). The phonon momentum of R mode is not a Moir\'{e} reciprocal vector of any commensurate superlattice, suggesting that the bilayer region is an incommensurate 30\textdegree-tBLG. It also reveals an unreported scattering process involving the reciprocal lattice vector of the bottom layer in this interesting 30\textdegree-tBLG region. This novel scattering process for the observed R mode is closely related to the interlayer interaction to be discussed below and is supported by the exotic electronic structure of the 30\textdegree-tBLG which we will investigate next.

\begin{figure}
	\subfloat{\label{fig3a}}
	\subfloat{\label{fig3b}}
	\subfloat{\label{fig3c}}
	\subfloat{\label{fig3d}}
	\subfloat{\label{fig3e}}
	\centering
	\includegraphics[width=7.3 cm] {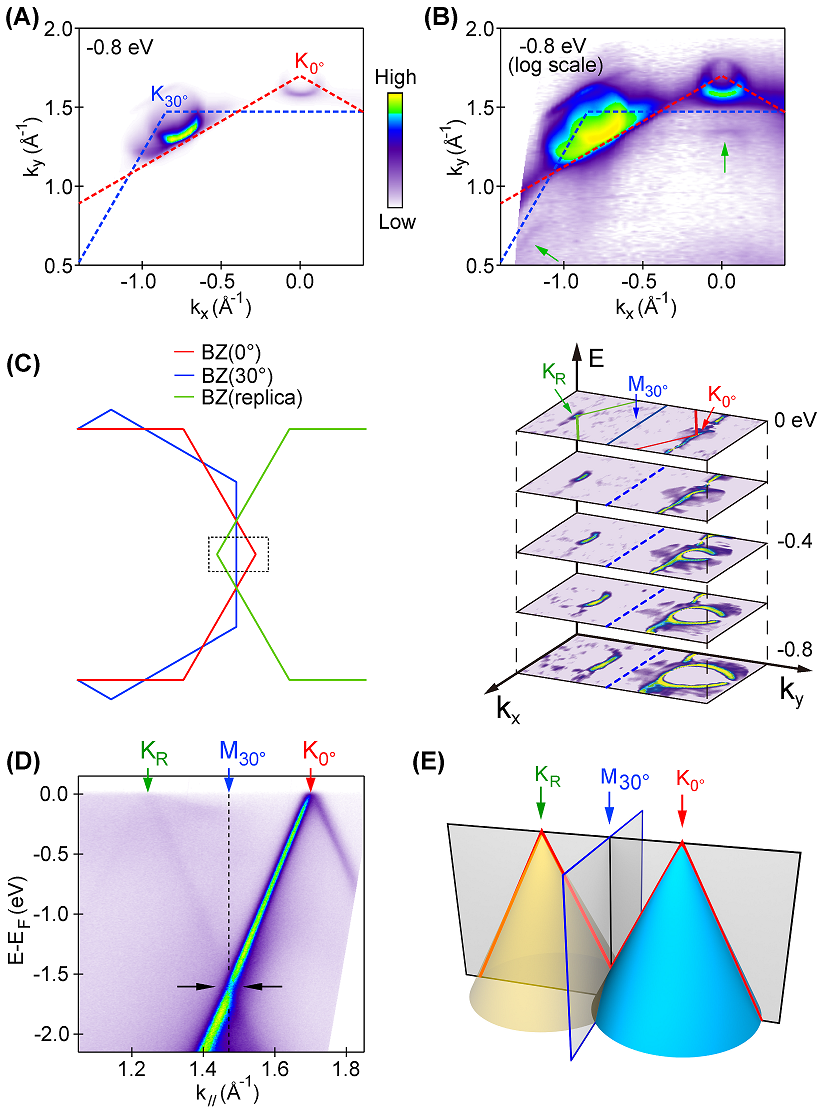}
	\caption{\textbf{Observation of mirrored Dirac cone in 30\textdegree-tBLG.}  \textbf{(A)}, Intensity map at -0.8 eV. The BZs of 0\textdegree-graphene and 30\textdegree-graphene are indicated by the dashed lines.  \textbf{(B)}, The same map as (A) but with logarithmic scale. Green arrows indicate the traces of replica Dirac cones. \textbf{(C)}, Schematic drawing of the Brillouin zone (left) and the 2D curvature image of constant energy maps for the region marked by rectangle at energies from E$_F$ to -0.8 eV (right). The BZ boundary of 0\textdegree, 30\textdegree~and replica bands are shown as red, blue and green lines.  \textbf{(D)}, The measured Dirac-type dispersion near the K point of the 0\textdegree~orientation. The 3 vertical arrows indicate the K point of 0\textdegree-graphene, the M point of 30\textdegree-graphene and the K point of the replica BZ. The gap is indicated by the horizontal arrows. \textbf{(E)}, Schematic illustration showing the mirrored relation between the original and the emerging replica Dirac cone.}
\end{figure}

\subsection*{Observation of mirrored Dirac cone}
The electronic structure of the sample was first measured by conventional ARPES with a beam size of $\sim$ 100 $\mu$m. Figure~\ref{fig3a} shows the ARPES intensity map at -0.8 eV. Characteristic conical contours appear at the K points of the graphene Brillouin zone (BZ) for both 0\textdegree~and 30\textdegree~layers (labeled as K$_{0^\circ}$ and K$_{30^\circ}$), in agreement with the coexistence of 0\textdegree- and 30\textdegree-graphene layers observed in the LEED pattern. Additional contours with weaker intensity are detected inside the first BZ near K$_{0^\circ}$ (green arrows in Fig.~\ref{fig3b}) when enhancing the weak features using logarithmic scale. The curvature images (Fig.~\ref{fig3c}) at different energies with higher visibility show that these contours expand from low to high binding energy similar to the Dirac cones at K$_{0^\circ}$, and they occur at the reflected position of K$_{0^\circ}$ (indicated by green arrow and labeled as K$_{\rm R}$) with respect to the 30\textdegree~BZ edges (blue lines).  The dispersion near the K$_{\rm R}$ also shows linear behavior, which is the reflected image of the Dirac cone at K$_{0^\circ}$ (Fig.~\ref{fig3d}). The mirrored Dirac cone shows similar intensity asymmetry caused by the dipole matrix element \cite{Shirley}, with stronger intensity between K$_{0^\circ}$ (K$_{\rm R}$) and the M point of the 30\textdegree~graphene layer M$_{30^\circ}$, suggesting that it comes from the original Dirac cone at the opposite momentum valley and is strongly related to the transfer mechanism discussed above for the double-resonance Raman mode (more details are shown in the Discussion section). Moreover, the suppression of intensity at the crossing point M$_{30^\circ}$ between the original and mirrored Dirac cones indicates a gap opening (pointed by black arrows in Fig.~\ref{fig3d}). Such mirrored Dirac cones and gap opening are not observed in the bottom monolayer 30\textdegree-graphene (see SI Appendix for details). Furthermore, the gap happens to appear at BZ boundary of bottom layer and is away from the substrate bands (see SI Appendix for details), suggesting that they are not caused by the graphene-substrate interaction, but intrinsic properties of the 30\textdegree-tBLG.  Therefore our ARPES results confirm that the 30\textdegree~graphene layer spatially overlap with 0\textdegree~graphene layer and they interact with each other. 

\begin{figure*}
	\subfloat{\label{fig4a}}
	\subfloat{\label{fig4b}}
	\subfloat{\label{fig4c}}
	\subfloat{\label{fig4d}}
	\centering
	\includegraphics[width=13 cm] {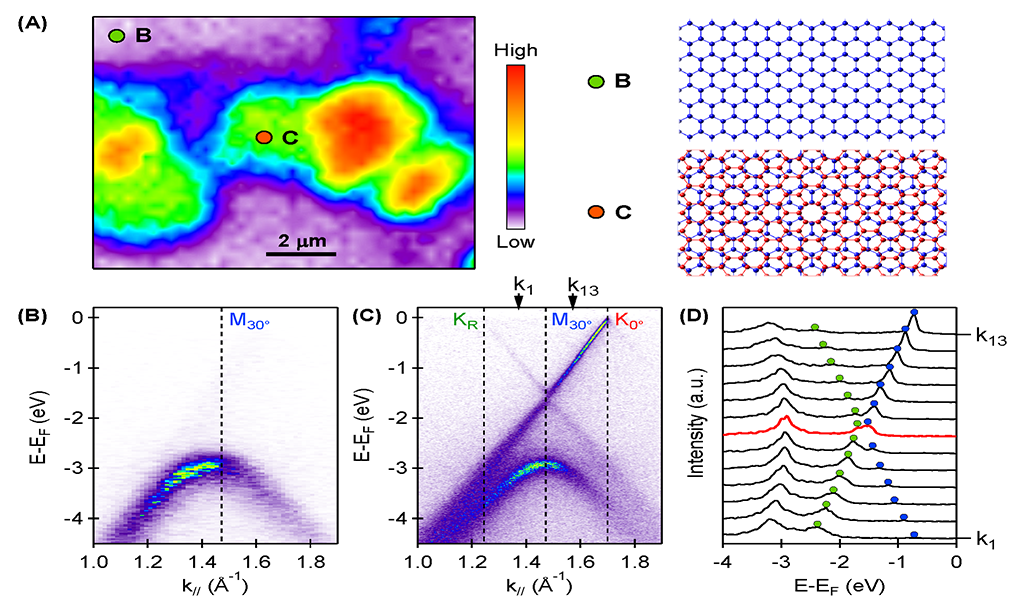}
	\caption{\textbf{Confirmation of 30\textdegree-tBLG and the electronic structure by NanoARPES measurements.}  \textbf{(A)}, Spatial intensity map obtained by integrating the energy from -1.5 eV to E$_F$. The structures for regions B and C are schematically shown on the right.  \textbf{(B), (C)}, Dispersion cuts taken at regions marked by \textbf{B} and  \textbf{C} in (A). \textbf{(D)}, The energy distribution curves (EDCs) taken in the momentum range marked by the arrows in (C). The EDC through the cross point of the original and mirrored bands is shown by the red color. Blue and green circles mark the peak positions.}
\end{figure*}

\subsection*{The band structures with spatial resolution}

The electronic structure of 30\textdegree-tBLG is further confirmed by NanoARPES \cite{NanoARPES}, which is capable of resolving the electronic band structure of coexisting graphene structures with spatial resolution at $\approx$ 120 nanometer scale. A spatially resolved intensity map is shown in Fig.~\ref{fig4a}, where we can distinguish different regions and map out their distinct electronic structure. Region labeled by \textbf{b} shows dispersion along the $\Gamma$-M direction of the 30\textdegree~bottom graphene layer (Fig.~\ref{fig4b}), while region \textbf{c} shows dispersions from  the 30\textdegree-tBLG, with dispersion along the $\Gamma$-K direction of 0\textdegree~top graphene layer coexisting with that along the $\Gamma$-M direction of the bottom 30\textdegree~graphene layer (Fig.~\ref{fig4c}). Therefore, dispersions measured by NanoARPES with spatial resolution provides definitive experimental evidence for the 30\textdegree-tBLG, and the mirrored Dirac cones indeed originate from this newly discovered tBLG structure. Moreover, by focusing only in the 30\textdegree-tBLG region, sharper dispersions can be obtained and the gap at the M point becomes more obvious. Analysis from the energy distribution curves (EDCs) for 30\textdegree-tBLG (Fig.~\ref{fig4d}) shows a gap at the M$_{30^\circ}$ point, suggesting the Dirac cone at K$_{0^\circ}$ and the mirrored Dirac cone are hybridized.

\begin{figure*}
	\subfloat{\label{fig5a}}
	\subfloat{\label{fig5b}}
	\subfloat{\label{fig5c}}
	\subfloat{\label{fig5d}}
	\subfloat{\label{fig5e}}
	\centering
	\includegraphics[width=15.5 cm] {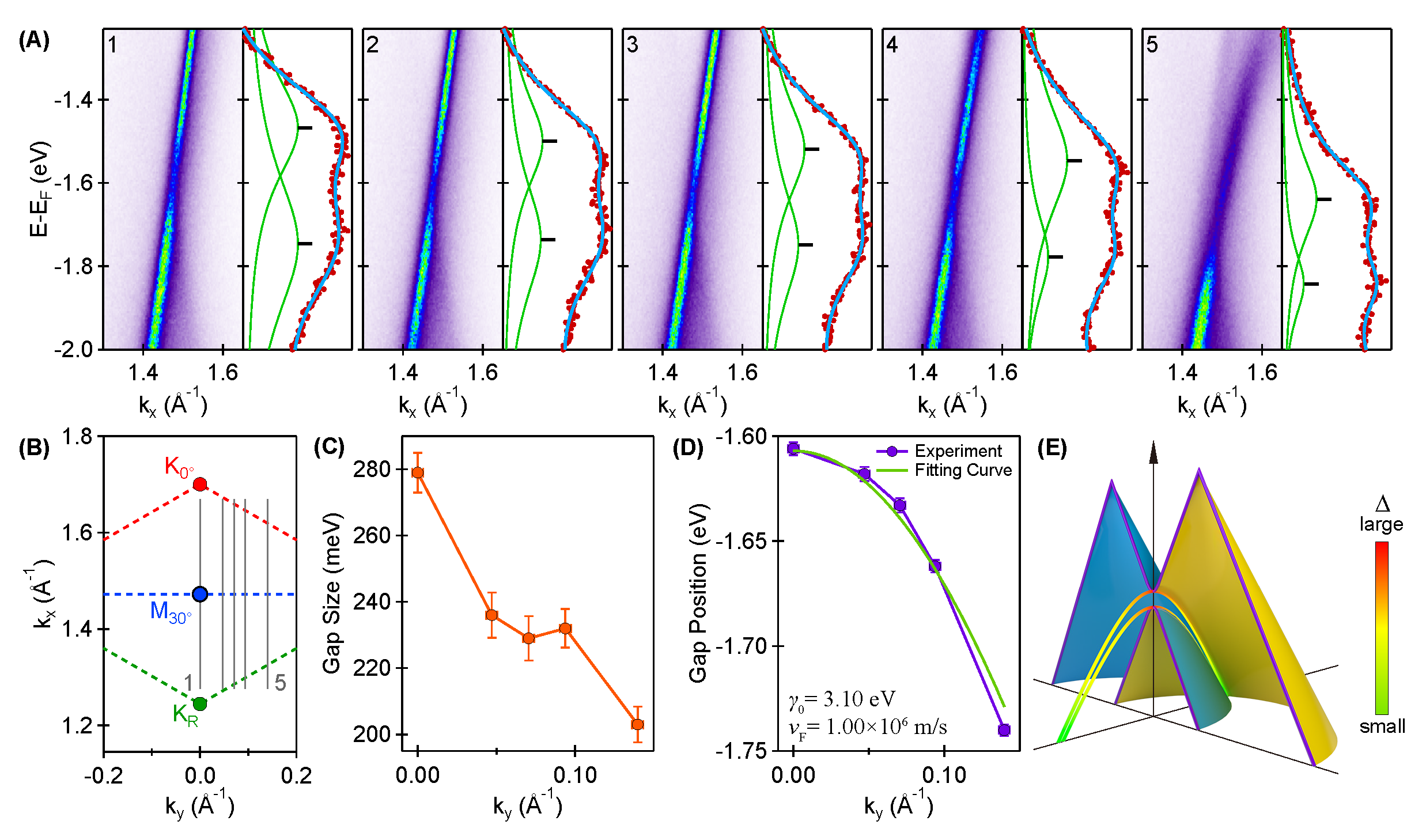}
	\caption{\textbf{Evolution of gap size and gap position in \textit{k}-space.}  \textbf{(A)}, 5 different cuts locating at different positions in the $k$-space shown in (B). The EDCs crossing the gap are shown on the right of the spectrum as red symbols. The fitting curves are plotted as blue lines and green curves are the two Lorentzian peaks from fitting results. Black markers show the positions of the two peaks. \textbf{(B)}, Schematic BZ showing the positions of the 5 cuts in (A). \textbf{(C)}, The gap size along the $k_y$ direction. \textbf{(D)}, The gap position along the $k_y$ direction. The green line is the fitting curve by the tight-binding model. \textbf{(E)}, Cartoon diagram for the two Dirac cones and their intersecting line with a gap. The color indicates the gap size.}
	\label{fig4}
\end{figure*}

\subsection*{Band gap in \textit{k}-space}
To track the evolution of the gap at the crossing points between the original Dirac cone at K$_{0^\circ}$ and the mirrored Dirac cone at K$_{\rm R}$, we show in Fig.~\ref{fig5a} dispersions measured from cut 1 to 5 (labeled in Fig.~\ref{fig5b}). The gap size is quantified by the peak separation of the EDCs at the crossing points and the extracted value is plotted in Fig.~\ref{fig5c}. It is clear that the gap size decreases from the maximum value of $\sim$ 280 meV at the M$_{30^\circ}$ point to $\sim$ 200 meV when deviating from the M$_{30^\circ}$ point. In addition, we can define the gap position as the average of the two peak positions. The gap position curve in Fig.~\ref{fig5d} shows that the gap is located at the intersecting line of the two Dirac cones, hence it is possible to fit the gap positions by the energy dispersion obtained from a tight-binding model (see SI Appendix for details). The fitting curve (green curve in Fig.~\ref{fig5d}) is in good agreement with the experimental data, giving a nearest-neighbor hopping parameter $\gamma_0$ to be 3.10 eV which is closed to the value derived from theoretical calculations \cite{Wallace1947}. The Fermi velocity $\nu_F=\sqrt{3}a\gamma_0/2$ ($a\sim$ 2.46 \AA) is extracted to be 1.003$\pm$0.002$\times10^6$ m/s. A schematic summary of the gap between the original and mirrored Dirac cones is shown in Fig.~\ref{fig5e}.

\begin{figure}
	\subfloat{\label{fig6a}}
	\subfloat{\label{fig6b}}
	\subfloat{\label{fig6c}}
	\centering
	\includegraphics[width=8.5 cm] {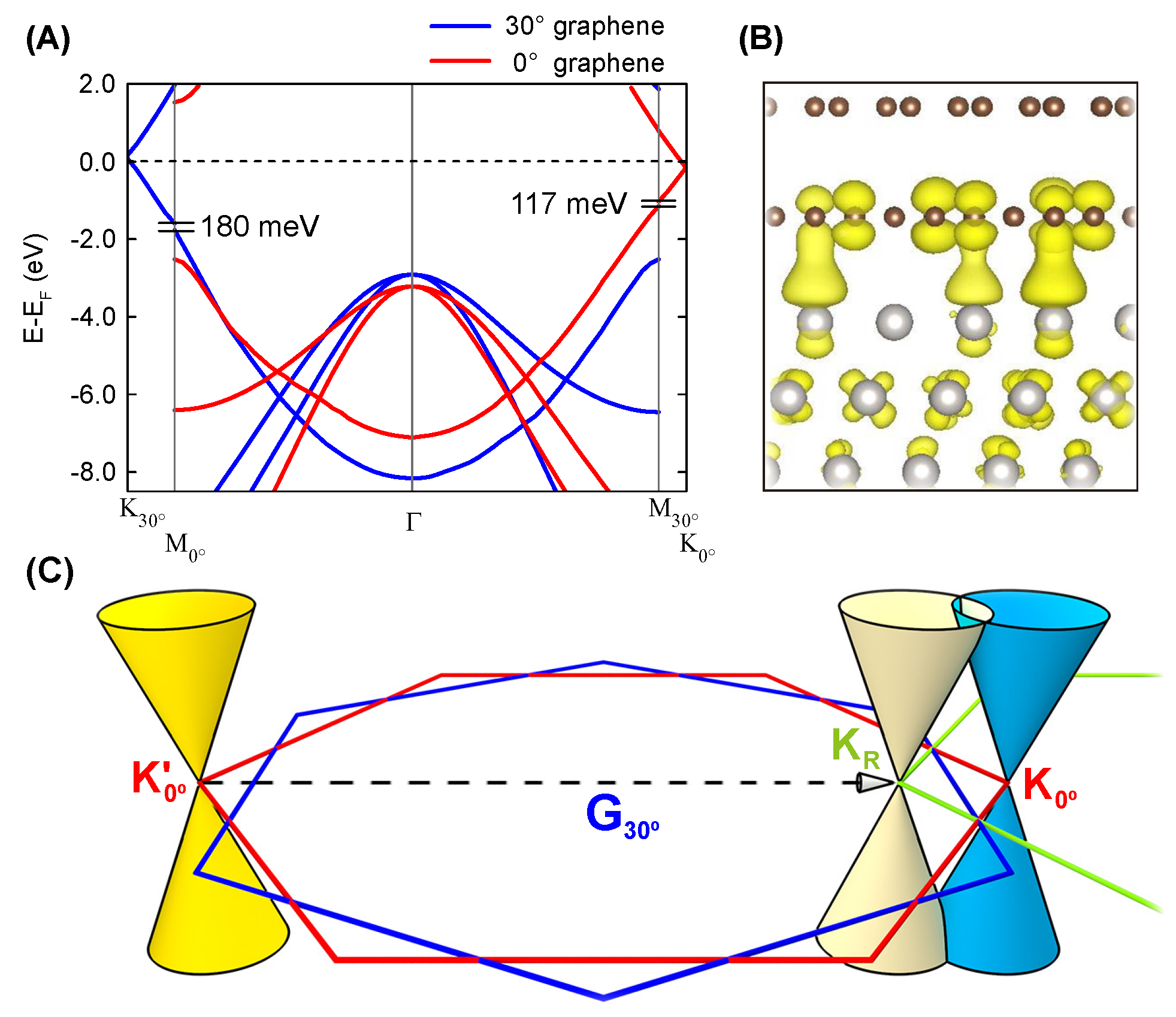}
	\caption{\textbf{Electronic structure of 30\textdegree-tBLG from first-principles calculations.}  \textbf{(A)}, Band structure of a $(5\times5)/(3\sqrt{3}\times3\sqrt{3})$R30\textdegree~tBLG in the extended zone scheme. The energy gap of $\pi$-band at M point of the other layer is indicated. \textbf{(B)}, Real-space projection of the Bloch state at $\Gamma$ point near Fermi level, showing a clear bonding state between $p_z$-orbital of C (brown balls) and $d$-orbital of Pt (white balls). \textbf{(C)}, Schematic illustration for the band structure of 30\textdegree-tBLG, showing the emergence of mirrored Dirac cones.}
	\label{fig6}
\end{figure}

\subsection*{Theoretical calculation}
\textit{Ab initio} calculations are performed to reveal the electronic structure of 30\textdegree-tBLG and investigate the stability. Since  Bloch theorem does not apply to the non-periodic structure and an infinitely large cell is beyond the calculation capability, a finite supercell of $(5\times5)/(3\sqrt{3}\times3\sqrt{3})$R30\textdegree~ is constructed to mimic the non-periodic 30\textdegree-tBLG system. The calculated electronic structure in Fig.~\ref{fig6a} shows band gap opening at the M point of the counterpart layer, with gap value close to the experimental measurements. An interesting observation regards the stability of 30\textdegree-tBLG when the Pt substrate is included in the calculation. While the formation energy of free-standing 30\textdegree-tBLG is 1.6 meV/(C atom) higher than the more common AB-stacking BLG, inclusion of the substrate leads to formation energy of 4.1 meV/(C atom) lower than AB-stacking one, which explains why 30\textdegree-tBLG emerges in our sample and suggests an important role of the Pt substrate in stabilizing the 30\textdegree-tBLG. By investigating real-space projection of the Bloch states, a significant coupling between the carbon p orbital and platinum d orbital (shown in Fig.~\ref{fig6b}) is revealed near Fermi level. Since such p-d coupling occurs at the $\Gamma$ point and phase matching between graphene and Pt substrate is less critical (due to zero $k$-vector). Therefore it could thus exist in systems with weak and incoherent sample-substrate interaction such as our sample, contributing positively to the stability of grown 30\textdegree-tBLG on Pt substrate. Extrapolating to larger supercell, like 5\texttimes5, 7\texttimes7 and 16\texttimes16, the stability of 30\textdegree-tBLG is further enhanced (See SI Appendix for details) and the electronic structure remains similar, suggesting that the calculation conclusion mentioned above still holds at the limit of infinitely large cell or the incommensurate structure. Our calculation, although not perfect, still provides some theoretical insights into the physics and stability of the 30\textdegree-tBLG.

\section*{Discussion}

Unlike the satellite Dirac cones appearing in commensurate graphene system with long-ranged Moir\'{e} pattern \cite{PletikosicPRL,Eli2011PRB,WEY2016A}, the emergence of mirrored Dirac cones in incommensurate 30\textdegree-tBLG indicates an unusual scattering mechanism in this novel structure. Based on a generalized Umklapp scattering process \cite{KoshinoNJP}, we build up a tight-binding model with second-order perturbation further included (See SI Appendix for details) and reveal a general coupling condition for two Bloch states in one layer of twisted bilayer system:
\begin{equation}
	\bm{k}_1^u = \bm{k}_2^u - \bm{G}^u + \bm{G}^d 
\label{eq2}
\end{equation}
where $\bm{k}_1^u$ and $\bm{k}_2^u$ are wave vectors of the two Bloch states in one layer, $\bm{G}^u$ and $\bm{G}^d$ are the reciprocal lattice vectors of the same layer and its counterpart respectively. Applying this condition, we propose a scattering mechanism for electrons in 30\textdegree-tBLG: Dirac cone at the K$^\prime$ point of one graphene layer K$_{0^\circ}^\prime$ is scattered by one reciprocal lattice vector of the other layer $\bm{G}_{30^\circ}$, forming a mirrored Dirac cone K$_R$ near the opposite momentum valley at K$_{0^\circ}$ as schematically shown in Fig.~\ref{fig6c}. This corresponds to the simplest case in \eqref{eq2} for $\bm{G}^u=\bm{0}$ and $\bm{G}^d=\bm{G}_{30^{\circ}}$. Such a mechanism is confirmed by three experimental observations. First of all, the momentum of the mirrored Dirac point $\bm{K}_{\rm R}$ is connected to the other graphene valley $\bm{K}_{0^{\circ}}^\prime$ by just one reciprocal lattice vector of the other 30\textdegree~layer $\bm{G}_{30^\circ}$. Secondly, the asymmetry of the intensity contour for the mirrored Dirac cone is identical to the one at K$^\prime$ (see Fig.~\ref{fig3d} or Fig.~\ref{fig4c}), suggesting that the mirrored Dirac cone is scattered from K$^\prime$ point. Thirdly, such a scattering mechanism is fully consistent with the observed R mode in Raman spectrum. Not only for the incommensurate structure, this scattering mechanism is also able to account for the appearance of replica Dirac cones in those commensurate graphene systems following the same analysis \cite{KoshinoNJP}. Thus, this mechanism can be applied to other van der Waals heterostructure beyond tBLG.

\section*{Conclusions}
In summary, we have successfully grown 30\textdegree-tBLG, a typical example for incommensurate superlattice with quasicrytalline order. The realization of 30\textdegree-tBLG provides new opportunities for investigating the intriguing physics of quasicrystalline superlattice.  Moreover, by revealing the mirrored Dirac cones in a 30\textdegree-tBLG, we provide direct experimental evidence for the strong interlayer coupling through a coherent scattering process in such an incommensurate superlattice. Such scattering mechanism can be applied to engineer the band structure of both commensurate and incommensurate tBLG as well as other van der Waals heterostructures.

\matmethods{

\subsection*{Sample Growth}
The graphene sample was obtained by annealing one Pt(111) substrate to about 1600 \textcelsius. At this high temperature, the carbon impurities would segregate from the bulk to surface forming graphene film.

\subsection*{ARPES and NanoARPES}
Conventional ARPES measurements were performed at the home laboratory with a Helium discharge lamp and beamline 10.0.1 of Advanced Light Source (ALS) with synchrotron radiation source. The NanoARPES measurements were performed at ANTARES endstation of Synchrotron SOLEIL, France, with a lateral spatial resolution of 120 nm. The sample temperature was kept at 20 K for measurements at ALS and 40 K for those at SOLEIL.

\subsection*{Calculations}
The calculations were based on VASP code \cite{VASP1,VASP2} with plane wave basis set \cite{PAW1,PAW2}. We adopted opt86b-vdW functional \cite{vdW1,vdW2} to include the van der Waals interactions. A periodic (5\texttimes5) on $(3\sqrt{3}$\texttimes$3\sqrt{3})$R30\textdegree~structure was created to simulate the 30\textdegree-tBLG because it is impossible to directly simulate the very large non-periodic bilayer. The lower layer is slightly compressed about 1.8\% to fit the Pt lattice and the upper layer is slightly stretched about 2\% to partially compensate the compression from the lower layer. Extensive convergence tests have been performed in terms of vacuum, film thickness, and k-point sampling. Although our calculations were based on one periodic structure, such effects would be extrapolated to non-periodic structures of 30\textdegree-tBLG.
}

\showmatmethods 

\acknow{This work is supported by the Ministry of Science and Technology of China (Grant No. 2016YFA0303004 and 2015CB921001), National Natural Science Foundation of China (Grant No.  11334006 and 11725418), Science Challenge Project (No. 20164500122) and Beijing Advanced Innovation Center for Future Chip (ICFC).
	
The Synchrotron SOLEIL is supported by the Centre National de la Recherche Scientifique (CNRS) and the Commissariat \`{a} l$^\prime$Energie Atomique et aux Energies Alternatives (CEA), France. This work is supported by a public grant overseen by the French National Research Agency (ANR) as part of the ``Investissements d$^\prime$Avenir'' program (Labex NanoSaclay, reference: ANR-10-LABX-0035), as well as the the French Ministre des affaires trangres et europennes (MAEE), the Centre National de la Recherche Scientifique (CNRS) through the ICT-ASIA programme grant 3226/DGM/ATT/RECH.}

\showacknow 


\bibliography{Ref_twBLG}

\end{document}